\documentclass[10pt,letterpaper,twocolumn]{article}

\usepackage[letterpaper,
  top=0.98in,
  bottom=0.98in,
  left=0.98in,
  right=0.98in,
  headheight=22pt,
  headsep=8pt,
  footskip=24pt]{geometry}

\usepackage[T1]{fontenc}
\usepackage{newtxtext,newtxmath} % Times-like text and mathematics
\usepackage{microtype}
\usepackage{graphicx}
\usepackage{booktabs}
\usepackage{array}
\usepackage{amsmath}
\usepackage{enumitem}
\usepackage{titlesec}
\usepackage{fancyhdr}
\usepackage{lastpage}
\usepackage{caption}
\usepackage{url}
\usepackage[hidelinks]{hyperref}

\usepackage[nolist]{acronym}

\usepackage{graphicx} % Required for inserting images

\newcommand{\IACTitle}{A Technosignature Survey Of The Anti-Solar Point: The SETI Institute’s Largest-Ever Campaign With The Allen Telescope Array}
\titleformat{\section}
  {\normalfont\bfseries\fontsize{10}{11}\selectfont}
  {\thesection.}{0.45em}{}
\titlespacing*{\section}{0pt}{9pt}{1pt}

\titleformat{\subsection}
  {\normalfont\itshape\fontsize{10}{11}\selectfont}
  {\thesubsection}{0.45em}{}
\titlespacing*{\subsection}{0pt}{7pt}{1pt}

\titleformat{\subsubsection}
  {\normalfont\itshape\fontsize{10}{11}\selectfont}
  {\thesubsubsection}{0.45em}{}
\titlespacing*{\subsubsection}{0pt}{6pt}{1pt}

\newcommand{\iacunnumbered}[1]{%
  \par\vspace{7pt}%
  \noindent\textbf{#1}\par\vspace{1pt}}

\setlist{nosep,leftmargin=*}

\begin{document}

% ============================================================
% Full-width title, author information, abstract, and keywords
% ============================================================
\twocolumn[{%
\begin{minipage}{\textwidth}
\begin{center}
  \fontsize{11}{12}\selectfont
  % \IACPaperCode\ % (Paper Code)

  \vspace{8pt}

  {\bfseries\fontsize{12}{14}\selectfont \IACTitle} % \(Title of Your Paper)

  \vspace{8pt}

  {\bfseries\fontsize{11}{13}\selectfont
  Isabel A. Gerrard\textsuperscript{a,b}, Sofia Z. Sheikh\textsuperscript{b,c}}
\end{center}

\vspace{5pt}

\noindent\textsuperscript{a}\textit{Breakthrough Listen, University of Oxford, Department of Physics, Oxford, UK, OX1 3RH
}\par
\vspace{-1pt}
\noindent\textsuperscript{b}\textit{The SETI Institute, Mountain View, CA, USA, 94043}\par
\vspace{-1pt}
\noindent\textsuperscript{c}\textit{Berkeley SETI Research Center, University of California, Berkeley, CA, USA, 94720}\par

\vspace{8pt}

% \noindent * Corresponding Author

\vspace{11pt}

\begin{center}
\textbf{Abstract}
\end{center}
\vspace{-4pt}

\noindent
The ATA-ASP campaign is a radio technosignature survey of the \acl{ASP} as a ``Schelling point'', using the \acl{ATA} in the \ac{SETI}. 
We borrow from game theory the concept of Schelling points—mutually determinable solutions between agents without communication—to hypothesize spacetime coordinates where both we and an extraterrestrial agent might send and search for a signal.
From the perspective opposite the Sun, once each year an extraterrestrial agent could detect Earth transiting within the habitable zone, thus this \acf{ASP} poses a shared and salient time-coordinate Schelling point. This campaign uses the \acf{ATA} to observe Earth’s transit when the Sun is directly opposite a potential transmitter in the sky (i.e. the ecliptic), and searches for sky-localized, narrowband, Doppler-drifting signals.
Observations use the \ac{ATA}’s 28 active 6.1 meter dishes tuned to 2356-3700 MHz in a two-beam beamformer mode for sky-localization. Analysis is conducted using the \acl{ATSAT} (\texttt{\acs{ATSAT}}), which integrates software packages \texttt{bliss} and \texttt{NBeamAnalysis} with \acl{RFI} removal and outputs spectrograms. \texttt{Bliss} identifies narrowband signals and their Doppler-drift rates to indicate technological and distant origin, while \texttt{NBeamAnalysis} compares  beams to localize signals to the \ac{ASP}. The plots are visually inspected and cataloged into six signal morphologies: one to categorize potential signals-of-interest and five for \acl{RFI}. 
To date, the ATA-ASP survey has observed approximately 12\% of the projected coordinate coverage over the course of 126 observations, 67\% of which have been analyzed, producing more than 11,000 plots. All plots have been inspected, and 75\% are labeled. Of the cataloged spectrograms, 25 contain narrowband drifting signals sky-localized to the ASP-beam, and the 10 corresponding \ac{ASP} coordinates are scheduled for re-observation.
Although \ac{SETI} literature has long proposed searching for beacons at the \ac{ASP}, this campaign provides the first observational coverage. Observing Earth’s ecliptic with the \ac{ASP} method will require a multi-year effort and will produce over 11 petabytes of data, making it the \ac{ATA}’s largest survey ever.

\vspace{2pt}
    \acf{ASP};
    Schelling point;
    \Acf{RFI};
    \acf{ATA}
\vspace{8pt}
\end{minipage}
}]

% ============================================================
% Main two-column manuscript
% ============================================================

\iacunnumbered{Acronyms/Abbreviations}
\acf{SETI}, \acf{ASP}, \acf{ATA}, \Acf{ETA}, \Acf{SI}, \acf{HCRO}, \acf{RFI}, \acf{SNR}, \acf{SOI}, \ac{ETZ}, \Acl{ATSAT} (\texttt{\acs{ATSAT}}), \Acl{BLADE} (\texttt{\acs{BLADE}}), System Effective Flux Density (SEFD)

\section{Introduction}
% The ATA-ASP campaign poses a comprehensive radio technosignature survey of the \acf{ASP}, and is the largest-ever survey conducted with the SETI Institute's \acf{ATA}.
% The survey reasons the \ac{ASP} could be a time-location coordinate, or ``Shelling point,'' where humanity and another intelligent civilization might both search for and direct interstellar communication.
This work builds upon the ATA-ASP survey begun in 2023 by Sheikh and Hort et al. for which the survey design and pilot results are currently in press (`the pilot paper') \cite{asp_pilot}. Henceforth, `we' refers to human civilization, as a technologically-advanced intelligent civilization mirroring an \acf{ETA}.

\subsection{Technosignatures and Schelling Points}\label{subsec:technosig_and_sp_into}
%%%%% technsignatures are X
In the \acf{SETI}, ``technosignatures'' refer to searching for non-human technology that serve as a proxy for detecting intelligent life complementary to ``biosignatures'' as signs of life \cite{jt_technosig}; a radio technosignature campaign searches for artificially-generated radio transmission as opposed to natural astronomical radio emission. 
%%%%% radio technosignature surveys are compelling because X
% \todo{Radio technosignatures, specifically, are compelling is a tdn{compelling, benefecial, good case for} interstellar communication because \tdn{we use for our comm (=normally radiated by tehcnical civ.s), req. least radiated power, efficient to generate and detect, travel at high speed, durable to absorportion and deflection}}
%%%%% 9 axes of permutation in radio searches 
% \tdn{Nine Axes of Merit for technosignature searches (Sheikh 2020)}
%%%%% search space is finite, but large
%%%%%  also how we reason SETI looks for beacons because
\ac{SETI} commonly assumes that an omnidirectional transmitter would be inefficient and unintended leakage radiation would be weak and difficult to detect, so instead searches for signals intentionally broadcast in specific directions, or ``beacons''.
%%%%% "cosmic haystack" %%%%%  where start, etner SP
Thus, to narrow the search in the vast ``cosmic haystack'' \cite{wolfe_haystack} \cite{wright_kanodia_lubar}, \ac{SETI} literature borrows the concept of ``Schelling points'' from game theory to postulate where an \acf{ETA} might direct such a beacon \cite{wright_2020}. 

%%%%%%%%%%%%%% SCHELLLING POINTS %%%%%%%%%%%%
A Schelling point is a default solution two agents both naturally arrive at, without communication, because of some shared cultural understanding of prominence in the given context \cite{schelling_1980}.
% For example, two friends find each other at the Big Ben at noon after planning to meet in London, but forgetting to arrange a meet-point.
The predictability of a Schelling point relies on a common understanding of the prominent features in this context, and the mutual awareness of each other's knowledge.

In the context of technosignatures, \ac{SETI} postulates spatiotemporal Schelling points \cite{davenport_2019} that both we and an \acf{ETA}, without established interstellar communication, could converge on sending and searching for a signal. Schelling point technosignature campaigns are hypotheses of the mutually-understood, salient information the survey's astronomical target is predicated upon. The underlying argument for its shared context hinges on using humankind as a point of reference for technologically-advanced intelligent life.

\subsection{The \Acl{ASP} as a Schelling point}
%%%%% what mutual knowledge we have with ETA 
%%%%% same sense we care about cataloguing exoplanets and their habitability (can sustain liquid water) 
%%%%% do so with transit method 
%%%%% salient mutual and self-referential (??) knowledge 
The argument behind the \acf{ASP} as a Schelling point hypothesizes that both we and an \ac{ETA} may catalogue habitable exoplanets using the transit method, then mutually-determine that this context provides a time and location to coordinate interstellar communication.

%%%%% do so with transit method 
The transit method measures the periodic dimming of a star's light to determine an orbiting planet crossing in front of it. An exoplanet is deemed ``habitable'', hence of possible interest to astrobiologists, by calculating its distance to the host star to establish if its surface temperature could sustain liquid water (i.e. ``the habitable zone'') \cite{kasting_1993}. Although this definition of habitability assumes that water is necessarily essential for life, it is practical to ground the search for life in life \textit{as we know it}. 

%%%%% ETA could detect us as such and know we are habitable
It follows that if an \ac{ETA} also used the transit method in their search for life \cite{singer1982look}, they would detect Earth against the Sun as habitable and classify Earth as an eligible target for interstellar communication \cite{corbet2003synchronized}.
%%%%% what is "ASP" specificcaly
Detecting habitable Earth using the transit method and a consequent effort to contact, requires the angle of the transmitting \ac{ETA} to be within range of the ecliptic (i.e. within the \acf{ETZ}) \cite{castellano_2004}.   
%%%%% salient mutual and self-referential (??) knowledge
%%%%% could inpsire both to converge on ASP as time-location for interstellar communication 
Therefore, in the Schelling point paradigm where we and an \ac{ETA} conclude the transit method could be a potentially shared concept in the search for life-context, the Earth's transit (i.e. ecliptic) is the prominent location feature.
In order to temporally synchronize finding and broadcasting a beacon within the \ac{ETZ}, a salient time might be when the Sun, the receiving Earth, and the transmitting \ac{ETA} align. The \acf{ASP} is the apparent position of this alignment from Earth, as when the Earth is at the midpoint of its transit from the \ac{ETA}'s perspective, the transmitter is directly opposite the Sun in the sky\footnote{This paradigm presupposes the transmitter accounts for light-travel time \cite{asp_pilot}.}. 

The \ac{ASP}, therefore, is a compelling astronomical context and spacetime coordinate backing a Schelling point. Accordingly, a comprehensive technosignature survey of every possible \ac{ASP} position necessitates a large-scale campaign to achieve observational coverage of the entire ecliptic.

\subsection{The \acl{ATA}}
The \acf{ATA} is a 42-element radio interferometer at the \acf{HCRO} located in Hat Creek, Northern California. The \ac{ATA} is currently operated by \acf{SI}, and was purpose-built for radio technosignature surveys \cite{welch2009}, making it especially suitable for the magnitude and objective of the ATA-ASP campaign. 
The array is sensitive from 1-10 GHz and up to four 672 MHz spectral windows (``tunings'') within this range can be observed simultaneously using the \ac{ATA}'s on-site local-oscillators. 
The \ac{ATA}'s 6.1 meter dishes are currently being retrofitted: 28 of 42 have been upgraded with cryocooled, log-pyramidal ``Antonio'' feeds making up the ``active array'' so far.
The active array also allows for up to two simultaneous beamformed beams as part of the high spectral resolution beamformer backend which runs on the \texttt{\ac{BLADE}} software \cite{blade_cruz_2024}.

\subsection{ATA-ASP Signals of Interest}\label{subsec:soi_intro}
The ATA-ASP survey searches for narrowband, Doppler-drifting signals, sky-localized to the \ac{ASP} (See `Sky-Localized SOI', Fig. \ref{fig:morphology_ex} in Appendix D). Together, these criteria suggest the signal is technological and distant in origin, as a proxy for intelligent and extra-terrestrial origin, thus characterizing a potential technosignature.

%%%%%%%% NARROWBAND
A narrowband transmission indicates that a signal may be technological in nature because a sufficiently narrow radio emission is distinct from the broader-band radio emission of natural astronomical phenomena, and narrowband radio transmission resembles our own communication technology \cite{cohen_1987}.
% Searching for a semblance to our communication technology is practical because, again, it grounds the search for technology in that as we know it--we can plausibly state that narrowband radio is normally radiated by technical civilizations \todo{cite}--and because the technical reasons that we developed narrow radio communication also justify its fitness for interstellar communication (e.g. travels at high speed, requires least radiated power, easy and efficient to generate and detect, durable to absorption and deflection by Earth's atmosphere and interstellar gas/dust, etc. \todo{cite}).

%%%%%%%% DRIFTING
A technosignature from a distant \ac{ETA} is also expected to drift in frequency over time due to the Doppler-effect, while locally transmitting \acf{RFI} would exhibit near-zero drift-rates.
% The \ac{ATA} does not perform any coherent de-Doppler methods to recover more power from drifting signals \cite{margot2021search}, but the \texttt{bliss} software within \ac{ATSAT} deploys some incoherent correction for drift rates (Section \ref{subsec:atsat_methods}). Though, it is also proposed a technologically advanced civilization could transmit signal \tdn{in a way to correct for Doppler-drift} \tdn{in attempt to convenience us} \todo{cite this as a theory}, the ATA-ASP survey does not mask near-zero drift-rates from the final output plots.

%%%%%%%% SKY-LOCALIZED
Finally, for a supposed beacon originating from a transmitter at the \ac{ASP}, a signal should be localized to the target \ac{ASP} coordinate on the sky (``sky-localized''). This manifests as the signal present only in the beamformed beam pointing at the \ac{ASP} (``ASP-beam''), or for the signal to be significantly attenuated away from the \ac{ASP} (``off-beam'')---quantified as the ``beam-ratio'' which compares the \acfp{SNR} between beams.

\subsection{The ATA-ASP Survey Size}\label{subsec:survey_size_intro}
Given that \acf{HCRO} is a ground-based facility, the \ac{ATA} can only observe the \ac{ASP} dependent on \ac{HCRO}'s latitude and longitude: when the array is opposite the sun (i.e. nighttime in Hat Creek) and when the \ac{ASP} is above the horizon. Provided that half of the ecliptic is inaccessible in a year (the portion that is daytime), the campaign requires at least two years of observing sessions to sample every possible \ac{ASP} along the ecliptic. Furthermore, due to the $\sim$365.25 day orbit of the Earth (inciting the leap day every 4 years), the accessible nighttime portion of the ecliptic slightly precesses in successive years, ergo total survey coverage requires staggering observations over the course of four years. Finally, note that the \ac{ATA}'s observation time is not solely dedicated to this survey, nor the \ac{HCRO}'s resources for the campaign's data storage and analysis, extending the survey's timeline in practice.

Since this hypothesis assumes an \ac{ETA} is timing their transmission to align with Earth mid-transit (i.e. on the bisector), total coverage of the ATA-ASP survey implies 100\%, or 360$^\circ$, in ecliptic longitude and 10\% in ecliptic latitude, because the sum of synthesized beams' diameters are 10\% that of the \ac{ASP} (Fig. \ref{app:asp_geometry} in Appendix E).
The complete coverage would consist of 87,660 individual, but overlapping, \ac{ASP} coordinates (``pointings'') (Appendix A). Each pointing is observed for five minutes (one ``scan''), and each scan is 125 GB, so surveying all 87,660 pointings would produce 11.25 PB of data. This is the minimum necessary data volume, but the multi-year campaign will exceed this lower bound due to re-observation of pointings. The survey will point at coordinates already observed both intentionally, to follow up on \ac{SOIs} or corrupted scans, or coincidentally (e.g. longer sessions in the wintertime will overlap in observed portions of the ecliptic between years).
% \todo{despite motivation for ASP techno. survey in literature, this ATA-ASP is first dedicated campaign to systematically and comprehensively cover ASP. other large radio technosignature surveys only include ASP positions fortuitiously/ not as inteded strategic target \todo{ex BLI, META, BETA and cite, cite, cite or don't mention}
% }
% -------------------- Materials and Methods --------------------
\section{Methods}

\subsection{Observing Strategy}\label{subsubsec:obs_strat}
The observing strategy follows that detailed in the pilot paper for the `Full ATA-ASP Survey' \cite{asp_pilot}.

The survey uses the \ac{ATA}'s 28-element active array in beamforming mode with two beams for sky-localization: the primary beam pointed at the location of the \ac{ASP} and the other a number of synthesized beamwidths away \cite{tusay_2024}, the ``ASP-beam'' and ``off-beam'', respectively. Each \ac{ASP} coordinate is observed for five minutes, with 1.9 Hz sampling and 16.8 second integration time, to sufficiently distinguish drifting signals in dynamic spectra and resolve Hz-scale narrowband signals without the \ac{ASP}'s position significantly precessing from the beamformed beam. 

The survey utilizes two consecutive 672 MHz tunings for a total 1,344 MHz bandwidth, with central frequencies at 2,692 and 3,364 MHz for continuous coverage from 2,356 to 3,700 MHz. This multi-year campaign maintains two tunings since its inception in 2023 \cite{asp_pilot}, though four local oscillators are operable as of Summer 2025. 
% \tdn{"Each night isn't a perfect series of mutually exclusive circles on the ecliptic, but they overlap (because the ASP doesn't move far in 5 min)" overlap in coverage of 80\% even at high frequnecies}

\begin{table}[htbp]
\caption{ATA-ASP Observation Parameters}
\label{tab:obs_params}
\centering
\fontsize{9}{10}\selectfont
\begin{tabular}{lr}
\toprule
\textbf{Parameter} & \textbf{Survey Value} \\
\midrule
Number of dishes & 28 \\
Central frequencies & 2,692 \& 3,364 MHZ\\
Observed range & 2,356--3,700 MHz \\
Total bandwidth & 1,344 MHz\\
Time integrations & 16.8 sec \\
Frequency resolution & 1.9 Hz \\
Scan length & 5 min \\
\bottomrule
\end{tabular}
\end{table}

\subsection{Data Analysis with ATSAT}\label{subsec:atsat_methods}
The data is analyzed using the \texttt{\acf{ATSAT}} pipeline: \acl{SI}'s efficient sky-localized, narrowband drifting signal search software\footnote{The software is available on GitHub \texttt{ATSAT} codebase: \url{https://github.com/SETIatHCRO/ATSAT} under an MIT license. The version used for this analysis is archived in Zenodo \cite{atsat_zenodo}.} \cite{atsat_zenodo}.
\texttt{\ac{ATSAT}} is a standalone Python package that integrates \texttt{bliss} \cite{bliss_zenodo} and \texttt{NBeamAnalysis} \cite{nbeam_zenodo} softwares within a custom architecture for \acf{RFI} excision, artifact filtering, and enhanced metadata logging. The pipeline outputs spectrograms of spatially-filtered hits with the greatest beam-ratios for visual inspection. The methodology for softwares \texttt{bliss} and \texttt{NBeamAnalysis} follows that detailed in the pilot paper \cite{asp_pilot}, though their application is now integrated into \texttt{\ac{ATSAT}} which the campaign used for all analysis beyond the pilot results.

The ATA-ASP survey utilizes \texttt{\ac{ATSAT}}'s ``frequency blanking'' to exclude hits in regions identified as \ac{RFI}-heavy in order to reduce downstream runtime and false positives in the output, a critical step to enable such a high-throughput campaign. For the majority of completed analysis to-date, the frequency blanking ranges sum to mask approximately 14.45\% of the total observed bandwidth (194.16 of 1,344 MHz) (Table \ref{tab:blanking_ranges} in Appendix C). However, it is significant to note that the frequency blanking ranges were periodically modified over the course of running analysis, since the long campaign timeline enables and requires iterative inspection of results and fine-tuning of parameters for future analysis (Fig. \ref{fig:signal_types_over_time} in Appendix F).  

\texttt{Bliss} detects Doppler-drifting, narrowband ``hits'', and \texttt{NBeamAnalysis} spatially compares hits to derive the ratio of the hits' \acl{SNR}s (SNRs) between beams (the ``beam-ratio''). \texttt{NBeamAnalysis} calculates the \ac{SNR}s, beam-ratio, and ``DOT score'' (dot-product) from a snippet of the dynamic spectrum from each beam, centered on the hit in the ASP-beam. In theory, the two snippets would be very different for a bright sky-localized signal, manifesting as a near-zero DOT score and a high beam-ratio between the ASP-beam versus the off-beam. To sky-localize a signal given a pair of spaced-apart synthesized beams, the ratio between the power in a coherent beam centered on an emitting target and an incoherent beam should be a factor of $\sqrt{N_{antennas}}$ in the worst case. Therefore, for the 28-dish active array, we would expect a beam-ratio of at least $\sqrt{N_{antennas}} = \sqrt{28} = \sim5.29$ for comparing two coherent beams. 
\texttt{\ac{ATSAT}} outputs any hits with beam-ratios greater than 5.29, capped at the 1,000 highest per observation, or else if none pass the threshold, the ``backup functionality'' plots up to 100 hits with the best combined score of high beam-ratio and low DOT score. 

\begin{table}[htbp]
\caption{ATA-ASP Analysis Parameters}
\label{tab:anal_params}
\centering
\fontsize{9}{10}\selectfont
\begin{tabular}{lr}
\toprule
\textbf{Parameter} & \textbf{Survey Value} \\
\midrule
Drift-rates searched by \texttt{bliss} & {$\pm$4 Hz/sec} \\
SNR threshold & 15 \\
SEFD & 272--291 Jy \\
Min. detectable flux density & 282--1,709 Jy \\
Total bandwidth blanked & 194.16 MHz\\
Observed bandwidth blanked & 14.45\% \\
Beam-ratio threshold & 5.29 \\
Output plots cutoff & 1,000 \\
Backup plots cutoff & 100 \\
% \todo{Minimum Detectable Flux at barycenter drift rate (W/m2) & {$3.9\times10^{−23}$} \\}
% \todo{Maximum Distance to detect a ∆νt = 1 Hz Arecibo-equivalent}
\bottomrule
\end{tabular}
\begin{flushleft}
The SEFD and minimal detectable flux density values were calculated in the pilot paper for the ``Full ATA-ASP survey'' \cite{asp_pilot}, though these values will likely change after calculating them for the first year of observations and eventually the entire survey. See the pilot paper for values and reasoning of the `minimum detectable flux at barycenter drift rate (W/m$^2$)' and the `maximum distance to detect a $\Delta\nu_t = 1$~Hz Arecibo-equivalent' as well.
\end{flushleft}
\end{table}

\subsection{Post-Processing}\label{subsec:pp_methods}
The output spectrograms are manually cataloged into six categories: four types of \acf{RFI}, one ``\acl{SOI}'' (\ac{SOI}) category, and one ``Other/Backup'' (Fig. \ref{fig:morphology_ex} in Appendix D). The \ac{RFI} types were identified from visual inspection of plots revealing distinct signal morphologies present in the output: ``horizontal lines'', ``fading signals'', ``block signals'', and narrowband drifting signals apparent in both beams. The \ac{SOI} category labels narrowband drifting signals sky-localized to the ASP-beam (Section \ref{subsec:soi_intro}) and is occasionally used to label unique plots for follow-up. ``Other/Backup'' groups all other non-SOI plots such as indiscriminate noise, a blank plot, or narrowband drifting signals (in either or both beams) which are below the beam-ratio threshold from the ``backup'' functionality (Section \ref{subsec:atsat_methods}).

\subsection{Divergence from Pilot Design}
The key differences between the ATA-ASP campaign described herein and the original pilot design are: the number of dishes in the array (formerly 20); observed bandwidth (formerly 1,064 to 2,378 MHz); fine channel width (formerly 1 Hz); time integrations (formerly 15 sec); excluding \texttt{turboSETI} from analysis; the development and application of \texttt{\acs{ATSAT}} for analysis; various iterations of frequency blanking ranges; cataloging  signal types of spectrograms; calculations of projected campaign size; and the extent of progress in observations and analysis \cite{asp_pilot}.

% -------------------- Results --------------------
\section{Discussion and Results}
The following results give campaign progress as of September 12th, 2026; observations and analysis are ongoing. These results account for ATA-ASP observations from August 30th, 2024 until August 25th, 2026.

\subsection{Observing Progress}
By progress on ASP-coordinate pointings out of a 360$^\circ$ ecliptic-total coverage, the campaign is approximately 12\% complete. The survey to-date has conducted 126 observing sessions including 131 nights, totaling $\sim$876 hours on sky and nearly 1.3 PB of data. It is to note that, on occasion, observations were left recording until the next night, for which calibration would hold, in so adding a night of pointings to a single session's data. 
\\
Table \ref{tab:obs_prog_results} provides different metrics of observational progress to-date. For each metric: `Total' gives the volume for the survey thus far, including repeated observations or corrupted instances of data that will be excluded from analysis and/or re-observed; `Quality' is the count remaining if those instances are excluded from `Total'; `Projected' is the minimum necessary coverage in a complete ATA-ASP campaign; `Progress' is `Quality' out of the `Projected'. Approximately 8\% of the data across observing sessions was corrupted or invalid in some way (e.g. lost delay engine, lost data transfer for part of the frequency band, observation ending early due to weather, etc.). The `Total' amount given for the `Data (TB)' metric already excludes corrupted instances when the data itself was lost. Campaign progress is quantified by completed pointings ($\sim$12\%) because all other metrics are subject to variation in how the multi-year campaign logistics will unfold in practice.

\begin{table}[htbp]
\caption{Calculating Observational Coverage Progress}
\label{tab:obs_prog_results}
\centering
\fontsize{9}{10}\selectfont
\begin{tabular}{lrrrr}
\toprule
\textbf{Metric} & \textbf{Total} & \textbf{Quality} & \textbf{Projected} & \textbf{$\sim$Progress}\\
\midrule
\# Nights & 131 & 121 & 730--1,460 & 8-17\% \\
\# Pointings & 11,313 & 10,511 & 87,660 & 12\%\\
Hr. on Sky & $\sim$876 & $\sim$806 & >7,305 & 11\% \\
Data (TB) & $\sim$1,290 & N/A & >11.25 & <10.6\% \\
\bottomrule
\end{tabular}
\end{table}

\begin{figure}[htbp]
  \centering
  \IfFileExists{figures/observed_so_far.png}{%
\includegraphics[width=0.90\columnwidth]{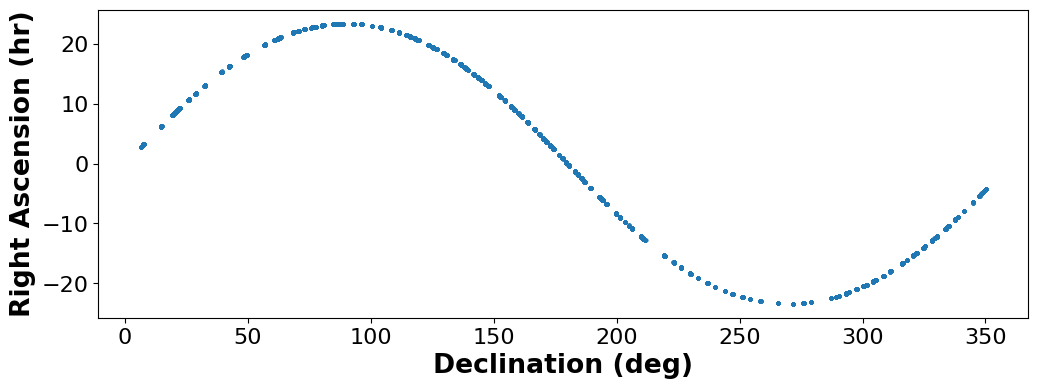}%
  }{%
    \fbox{\rule{0pt}{1.35in}\rule{0.90\columnwidth}{0pt}}%
  }
  \caption{Observed points in the ATA-ASP survey from 2024 to 2026: 11,313 pointings, including repeats.}
  \label{fig:pointings_so_far}
\end{figure}

Figure \ref{fig:pointings_so_far} plots the observed pointings on the ecliptic to show progress out of all possible ASP coordinates.

\subsection{Analysis Progress}\label{subsec:anal_prog}
66.67\% of the campaign has been analyzed: \texttt{\ac{ATSAT}} runs on a single observing session at a time and has completed 84 of 126 sessions.
Of the analyzed sessions, \texttt{bliss} found nearly 170 million narrowband drifting hits, frequency blanking removed approximately 55.65\%, and the spatial filtering step by \texttt{NBeamAnalysis} screened an additional $\sim$63.68\% hits. Of the remaining $\sim$27 million hits processed by \texttt{NBeamAnalysis}, over 11,000 were ultimately selected for plotting. All spectrograms have been inspected and about 75\% labeled (Section \ref{subsec:labeled_results}). Crucially, these $\sim$11,000 plotted hits include sessions analyzed early in campaign analysis before iterative inspection of \ac{RFI} in the outputs enabled gradual fine-tuning of frequency blanking ranges. Consequently, earlier sessions reached the 1,000 plotting-cap more often (Section \ref{subsec:atsat_methods}), and the number of spectrograms output per session will continuously reduce while the campaign progresses, due to optimizing \texttt{\ac{ATSAT}}'s \ac{RFI} excision from concurring output inspection. 

\begin{table}[htbp]
\caption{Volume of Hits Reduced and Processed by each \texttt{\ac{ATSAT}} Filtering Step}
\label{tab:dataload_so_far}
\centering
\fontsize{9}{10}\selectfont
\begin{tabular}{lrr}
\toprule
\textbf{Analysis Step} & \textbf{Hit Count} & \textbf{$\sim$Reduction} \\
\midrule
Found by \texttt{bliss} & 168,721,405 & N/A \\
After frequency blanking & 74,834,124 & 55.65\% \\
After spatial filtering & 27,178,030 & 63.68\% \\
Plotted & 11,212 & 0.04\% \\
Cataloged & 8,473 & N/A\\
\bottomrule
\end{tabular}
\end{table}

\subsection{Types of Signals in ATA-ASP Survey Outputs}\label{subsec:labeled_results}
%%%% 94% = 79/84 sessions analyzed, 75% = 8473/11212 plots
Approximately 94\% of analyzed sessions, or 75\% of the output spectrograms, have been cataloged into six \acf{RFI} and \acf{SOI} signal classes: ``horizontal line,'' ``fading signal,'' ``block signal,'' ``narrowband drifting in both beams'', ``Sky-localized \ac{SOI}'' for narrowband drifting signals brighter in the ASP-beam or other unique plots for followup, and ``Other/Backup'' for morphologies not-otherwise-specified and narrowband drifting signals below the beam-ratio threshold (Sections \ref{subsec:atsat_methods}, \ref{subsec:pp_methods}, and Fig. \ref{fig:morphology_ex} in Appendix D). 

Of the 8,473 cataloged plots,
    46.02\% are horizontal lines,
    43.08\% are fading signals,
    1.86\% are block signals,
    and 8.13\% are other morphologies or below beam-ratio threshold. 

Less than 1\% of all cataloged plots depict the two classes of narrowband drifting signals that \texttt{NBeamAnalysis} marks as surpassing the beam-ratio threshold: 0.61\% have those apparent in both beams, and the sky-localized \ac{SOI} class is the smallest category with 0.29\% of all outputs. There were 25 total \acf{SOIs}--spectrograms depicting narrowband drifting signals brighter only in the ASP-beam. These 25 sky-localized \ac{SOIs} span 10 independent frequencies and 10 individual ASP coordinates, their pointings observed across 7 different nights. One \ac{SOI} central frequency is repeated between two different ASP pointings within $\sim$0.13 MHz of each other. 14 of the 25 \ac{SOIs} are from a single pointing and session, all with central frequencies within $\sim$0.1 MHz (Fig. \ref{fig:signals_cats_by_freq} in Appendix G).
% \todo{insert signal type distribution}

% \todo{8473/11212\% plots} have been 
% Of all survey output spectrogram, \todo{8,241} plots fall into the morphology-specified classes \tdn{as 1-5 aka not-other/Backup}, the majority of \ac{RFI} present in the outputs appear horizontal lines or fading signals (\dc{46.02\% }and \dc{43.08\%} respectively), with only \dc{1.86\%} block signal. 

\begin{table}[htbp]
\caption{Signals Types of Cataloged Output}
\label{tab:category_distr}
\centering
\fontsize{9}{10}\selectfont
\begin{tabular}{lr}
\toprule
\textbf{Category} & \textbf{Count} \\
\midrule
Horizontal Line & 3,877 \\
Fading Signal & 3,629 \\
Block Signal & 157 \\
Narrowband, Drifting Both Beams & 51 \\
\textbf{*}Sky-localized, Narrowband, Drifting & 25 \\
Other/Backup & 685 \\
\midrule
Total Cataloged & 8,473 \\
\bottomrule
\end{tabular}
\vspace{2pt}
\begin{flushleft}
\textbf{*} \Acf{SOI}
\end{flushleft}
\end{table}

% -------------------- Discussion and Conclusions --------------------
\section{Conclusions}
This campaign poses the \acf{ASP} when Earth is mid-transit as a Schelling point to spatially and temporally synchronize interstellar communication and to narrow the \acf{SETI}. Though existing \ac{SETI} literature has proposed the \ac{ASP} as a compelling Schelling point, the ATA-ASP survey arranges the first systematic plan for complete 360-degree coverage of the ecliptic and the first observational coverage. The survey will require over four years of observing, more than 11 petabytes of data, and upwards of 700 recording sessions, making the ATA-ASP campaign the largest-ever campaign with the \acl{ATA}. Finally, incorporating Schelling points from game theory as a socio-psychologically informed heuristic demonstrates the value of cross-disciplinary approaches in exploratory sciences such as \ac{SETI} and in tackling the vast ``cosmic haystack''.

\subsection{Future Work}\label{subsec:future_work}
The corresponding \ac{ASP} coordinates for all sky-localized, narrowband Doppler-drifting signals have been, and will be, scheduled for re-observation to investigate \acf{SOIs}. Depending on inaccessibility of the ecliptic in a given year, \ac{SOI} pointings may be additionally, preemptively observed outside of their specific \ac{ASP} timing. All observations with any data corruption will be scheduled for re-observation to ensure total ASP coverage quantified by possible pointings. Future observations will be planned according to coordinates accessible on the ecliptic from the \acl{HCRO}, whereby coverage of possible pointings is maximized and overlap minimized (Section \ref{subsec:survey_size_intro}, Fig. \ref{fig:points_nxt_year} in Appendix B).

As data analysis progresses, output plots and metadata are continuously reviewed and analyzed to implement more sophisticated \ac{RFI} mitigation in an effort to extract increasingly interesting signals. Furthermore, the scale of the ATA-ASP survey and the novel additional cataloging process have created a large labeled dataset of radio spectrograms that will enable future work in \ac{RFI} characterization and the potential to train more advanced classification models. 

The original ATA-ASP survey design and pilot results are in press at \textit{AAS Journals} \cite{asp_pilot}. A paper documenting one full year (2025) of completed observations and analysis will begin preparation shortly thereafter. This followup paper of 2025-coverage will also resolve the complication of various frequency blanking ranges applied at different times convoluting analysis in post-processing (Section \ref{subsec:atsat_methods}, Table \ref{tab:blanking_ranges} in Appendix C, Fig. \ref{fig:signals_cats_by_freq} in Appendix E). The 2025 followup will use the cataloged outputs to precisely characterize \acl{RFI} in the ATA-ASP environment to determine optimal frequency blanking, then will re-analyze all observations post-\texttt{bliss} with uniform parameters and potentially with masking morphologies dynamically (i.e. not blanking entire ranges across every scan) (Fig. \ref{fig:signals_cats_by_freq} in Appendix G). 

% -------------------- Acknowledgements --------------------
\iacunnumbered{Acknowledgments}
The \acf{ATA} refurbishment program and its ongoing operations have received substantial support from Franklin Antonio. Additional contributions from Frank Levinson, Greg Papadopoulos, Breakthrough Listen Initiative and other private donors have been instrumental in the renewal of the \ac{ATA}. Breakthrough Listen is managed by the Breakthrough Initiatives, sponsored by the Breakthrough Prize Foundation. The Paul G. Allen Family Foundation provided major support for the design and construction of the \ac{ATA}, alongside contributions from Nathan Myhrvold, Xilinx Corporation, Sun Microsystems, and other private donors. The \ac{ATA} has also been supported by contributions from the US Naval Observatory and the US National Science Foundation.

% -------------------- Appendix --------------------
\iacunnumbered{Appendix A: Total Number of ATA-ASP Pointings}\label{app:calculate_pointings}
The 87,660 total ATA-ASP pointings in the survey ($N_{total}=87,660^\circ$) is based on the total coverage being the entire ecliptic (Section \ref{subsec:survey_size_intro}), so is given by the product of the individual ATA-ASP pointings per degree ($M_{pointings}=243.5$)(Eq. \ref{eq:b}) and the degrees in the entire ecliptic ($E_{deg}=360^\circ$):
\begin{equation}
\label{eq:n}   
       N_{total} = M_{pointings} \times E_{deg}
\end{equation}
where \\
\indent\(N_{total}\) is Pointings in ecliptic = 87,660\\
\indent\(M_{pointings}\) is Pointings in degree of ecliptic = 243.5\\
\indent\(E_{deg}\) is Degrees in ecliptic = 360$^\circ$
\\
The number of individual ATA-ASP pointings per degree $M_{pointings}$, is given by the reciprocal (Eq. \ref{eq:b}) of how many degrees the ASP moves in-between ATA pointings $M_{deg}$ (Eq. \ref{eq:a}):
\begin{equation}
\label{eq:a}  
    M_{deg} = E_{deg} \div d \div m \times s
\end{equation}
where \\
\indent\(d\) is Days per year = 365.25 days\\
\indent\(m\) is Minutes per year = $60\times24=1,440$ min\\
\indent\(s\) is Minutes per pointing = 6 min\\
\indent\(M_{deg}\) is Degrees per pointing
\begin{equation}
\label{eq:b}  
    M_{pointings} = (M_{deg})^{-1}
\end{equation}
There are 6 minutes between each pointing ($s=6$) as we record one 5-minute scan per ASP coordinate (Section \ref{subsubsec:obs_strat}) plus 1 minute encompassing data transfer, the backend restarting, and the instruments slewing. 

\iacunnumbered{Appendix B: Observable ASP coordinates in the next year}\label{app:2627_pointings}
\begin{figure}[htbp]
  \centering
  \textbf{Observable ASP coordinates in the next year}
  \IfFileExists{figures/already_observed_example.png}{%
    \includegraphics[width=0.90\linewidth]{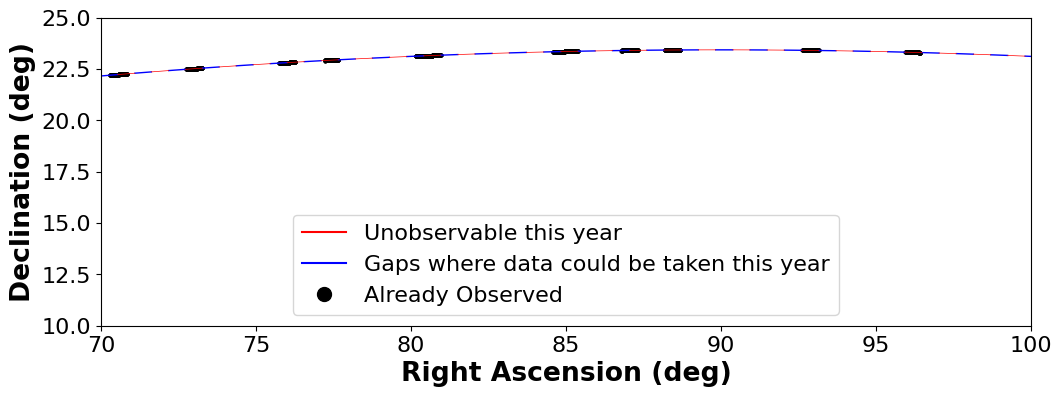}%
  }{%
    \fbox{\rule{0pt}{1.35in}\rule{0.90\linewidth}{0pt}}%
  }
  \caption{A portion of the 2026--2027 sky filtered by whether it has already been observed. The blue gaps are ASP coordinates we can aim to observe in the coming year that are accessible on the ecliptic (not-red) and have not already been observed (black).}
  \label{fig:points_nxt_year}
\end{figure}

\iacunnumbered{Appendix C: Current Frequency-Blanking Ranges}\label{app:blanking_ranges}
\begin{table}[htbp]
\caption{Current Frequency-Blanking Ranges}
\label{tab:blanking_ranges}
\centering
\fontsize{9}{10}\selectfont
\begin{tabular}{|l|l|l|}
\hline
2355.75--2359.50 & 3264--3274 & 3618.5--3619.0 \\
2480--2500 & 3341--3343 & 3640.18--3640.58 \\
2565.75--2585.00 & 3492.75--3492.85 & 3666.8--3667.0 \\
2590--2609 &3657.50--3657.65 & 3668.07--3668.7 \\
2705--2800 & 3615.4--3615.8 & 3675.04--3675.36 \\
3251.8--3253.0 & 3616.15--3616.60 & 3683.44--3684.25 \\
\hline
\end{tabular}
\end{table}
The frequency ranges identified as \ac{RFI}-heavy regions to blank hits in \texttt{\ac{ATSAT}}. These are the most up-to-date ranges used in current analysis, though it's significant to note the blanking ranges have changed over the course of analysis.

\iacunnumbered{Appendix D: Examples of Cataloged Signal Morphology Types}\label{app:morphology_ex}
\begin{table}[htbp]
\caption{Examples of Cataloged Signal Morphology Types}
\label{fig:morphology_ex}
\centering
\fontsize{9}{10}\selectfont
\begin{tabular}{c}
% \toprule
\textbf{Horizontal Line} \\
\includegraphics[width=0.90\columnwidth]{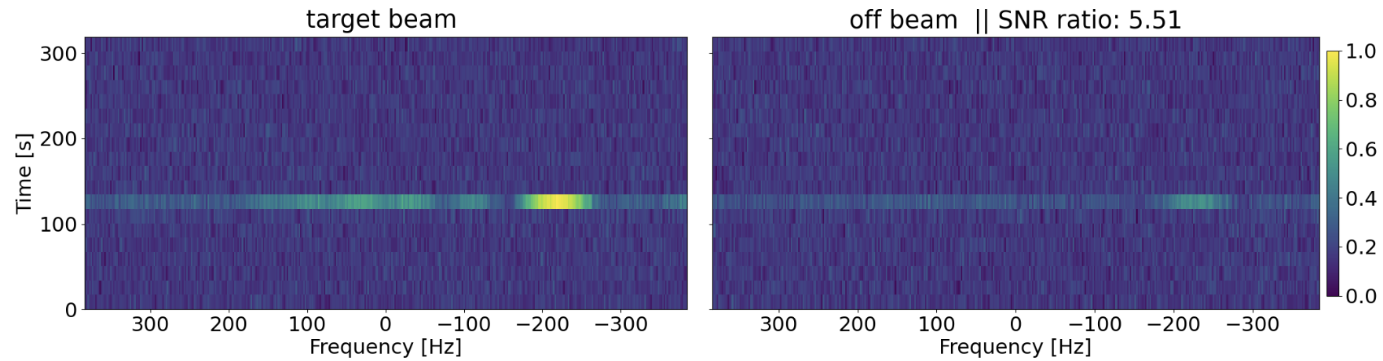} \\
% \midrule
\textbf{Fading Signal} \\
\includegraphics[width=0.90\columnwidth]{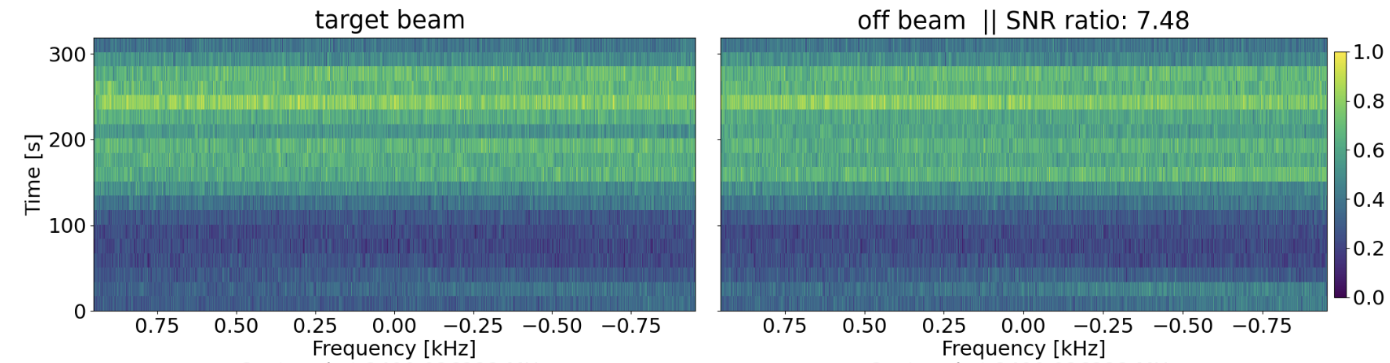} \\
% \midrule
\textbf{Block Signal} \\
\includegraphics[width=0.90\columnwidth]{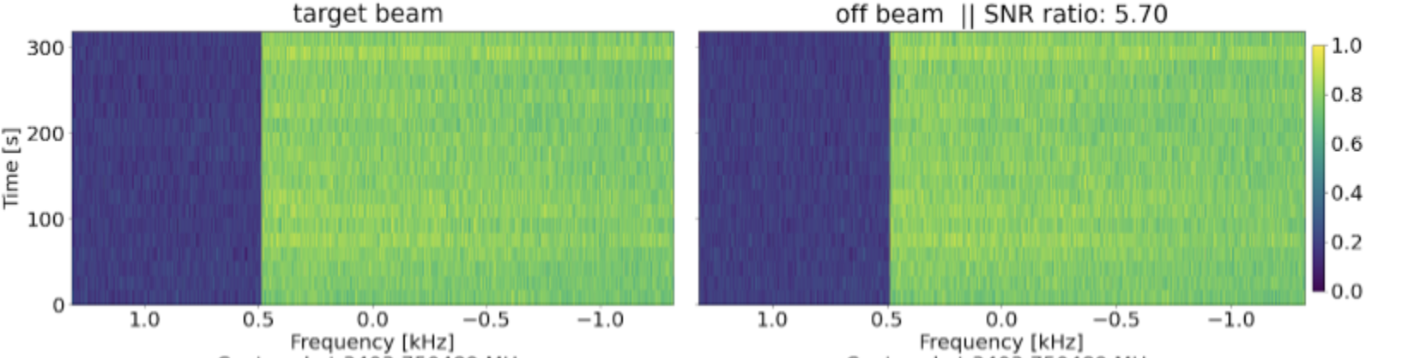} \\
% \midrule
\textbf{Narrowband Drifting Both Beams} \\
\includegraphics[width=0.90\columnwidth]{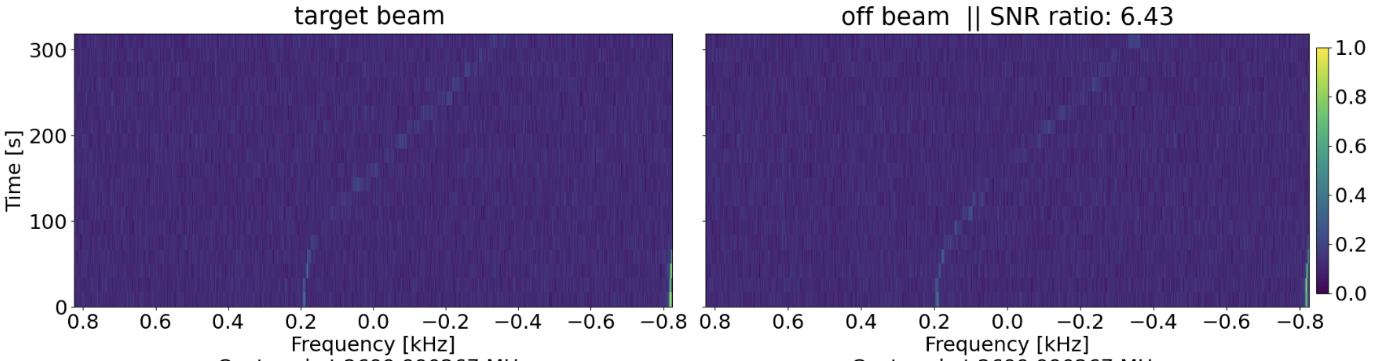} \\
% \midrule
\textbf{Sky-Localized SOI} \\
\includegraphics[width=0.90\columnwidth]{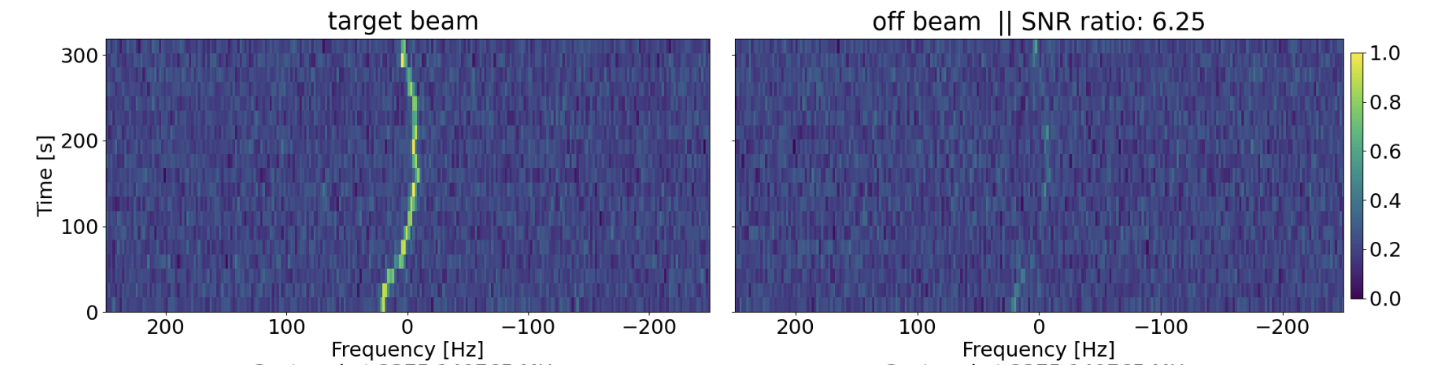} \\
% \bottomrule
\end{tabular}
\end{table}

\iacunnumbered{Appendix E: ATA-ASP On-Sky Geometry}\label{app:asp_geometry}
\begin{figure*}[htbp]
  \centering
  \textbf{ATA-ASP On-Sky Geometry}\par\medskip
  \IfFileExists{figures/categories_per_freqs.png}{%
    \includegraphics[width=\linewidth]{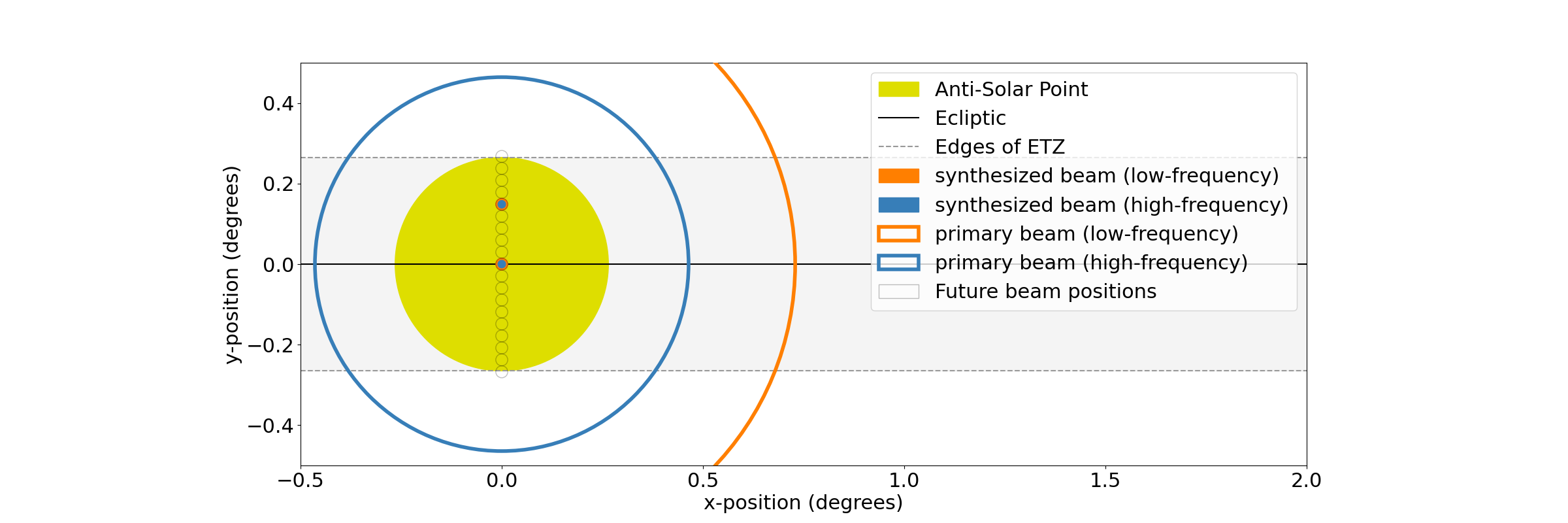}%
  }{%
    \fbox{\rule{0pt}{1.35in}\rule{\linewidth}{0pt}}%
  }
  \caption{The on-sky geometry of the \acf{ASP}, the ecliptic, and the synthesized beams of the ATA-ASP campaign. See the pilot paper by Sheikh and Hort et al. 2026 for full detail \cite{asp_pilot}.}
  \label{fig:asp_geometry}
\end{figure*}

\iacunnumbered{Appendix F: Distribution of Signal Types in Outputs over Time of Observation}\label{app:signal_types_over_time}
\begin{figure*}[htbp]
  \centering
  \textbf{Distribution of Signal Types in Outputs over Time of Observation}\par\medskip
  \IfFileExists{figures/stacked_types_over_time.png}{%
    \includegraphics[width=0.90\textwidth]
    {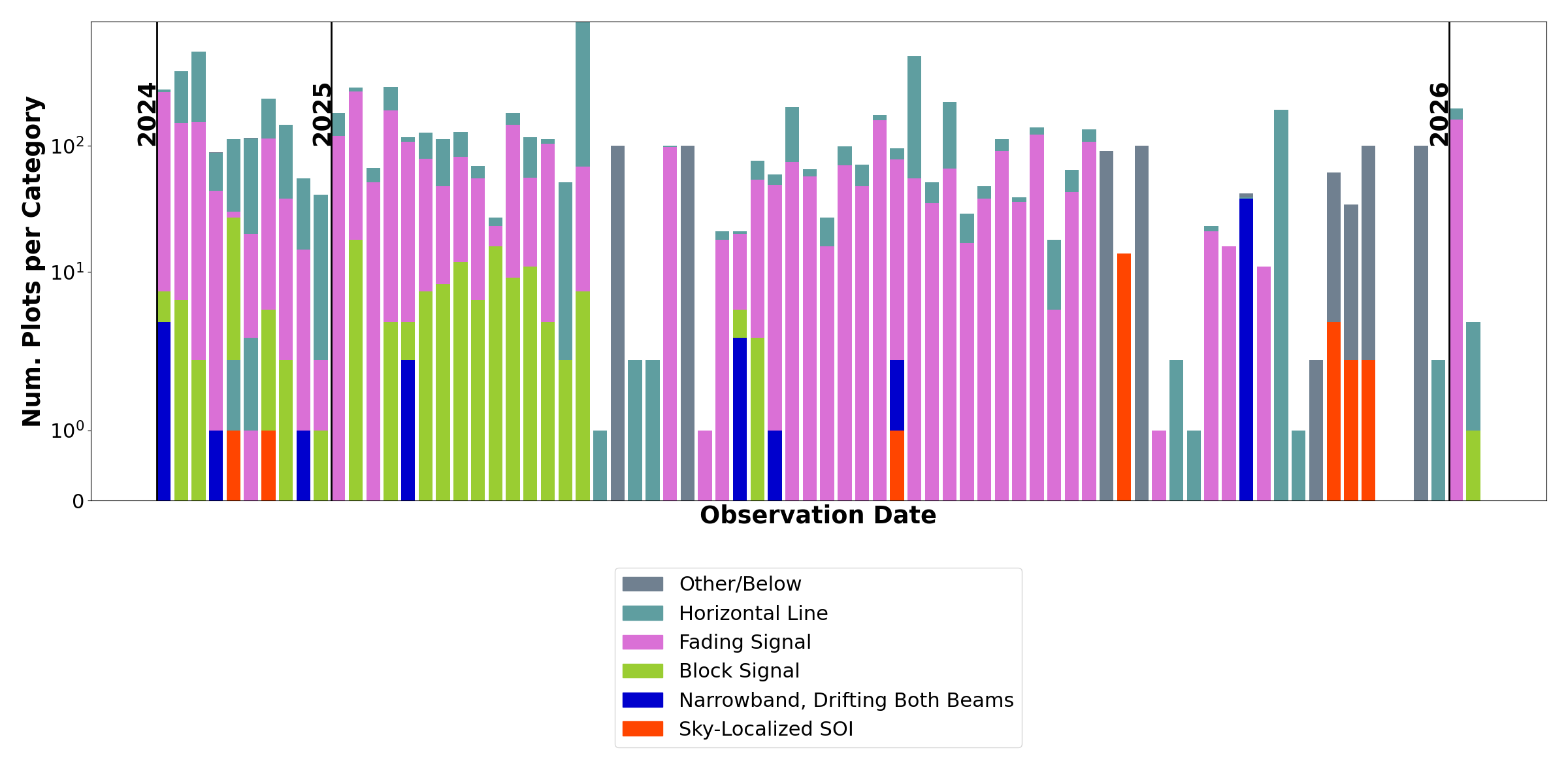}%
  }{%
    \fbox{\rule{0pt}{1.35in}\rule{0.90\textwidth}{0pt}}%
  }
  \caption{
  Evaluating how the type-distribution varies with time of observation is significantly convoluted by having sporadically modified frequency blanking ranges over the course of analysis. Furthermore, sequential observations were not always analyzed chronologically. This complication is a necessary product of the scale of the survey: iterations of inspecting outputs and adjusting blanking ranges was essential to streamline analysis of such a large dataset and to screen false negatives from plots that must be manually inspected. For example, often an \ac{RFI}-heavy region was only evident after seeing the final outputs, then the respective range was added/adjusted, to avoid it contaminating future outputs. Therefore, incrementally fine-tuning RFI excision also gradually decreases the overall number of plots output over the course of campaign analysis. Future dataset releases will employ a uniform set of blanking ranges across observations to remove these time-varying effects due to fine-tuning (Section \ref{subsec:future_work})
  }
  \label{fig:signal_types_over_time}
\end{figure*}

\iacunnumbered{Appendix G: Distribution of Signal Morphologies by Frequency}
\begin{figure*}[htbp]
  \centering
  \textbf{Distribution of Signal Morphologies by Frequency}\par\medskip
  \IfFileExists{figures/categories_per_freqs.png}{%
    \includegraphics[width=0.90\textwidth]{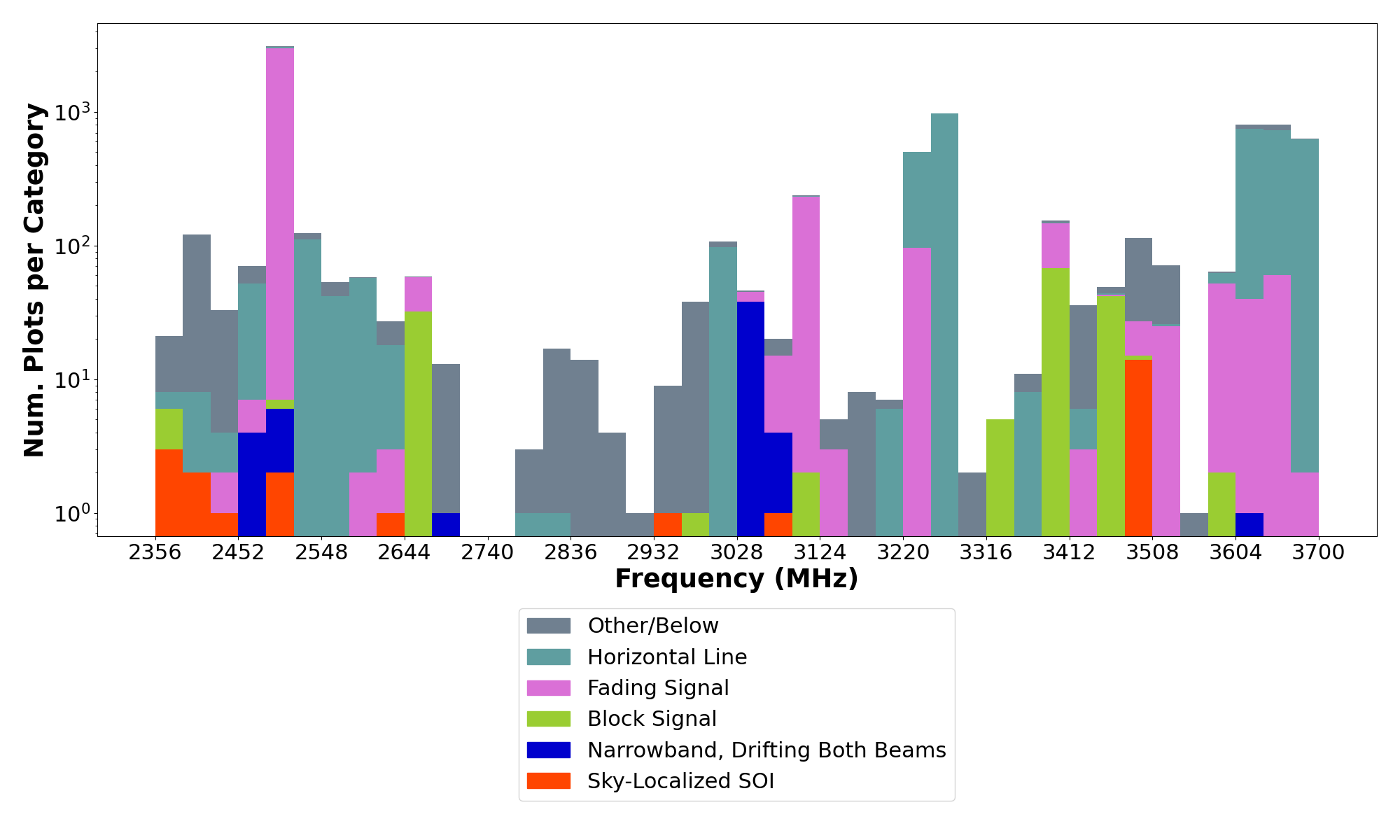}%
  }{%
    \fbox{\rule{0pt}{1.35in}\rule{0.90\textwidth}{0pt}}%
  }
  \caption{Frequency distribution of all cataloged hits in survey, stacked by signal type. This highlights \ac{RFI}-dense frequencies specific to underlying signal morphology, facilitiating fine-tuning frequency blanking ranges or potential dynamic masking.}
  \label{fig:signals_cats_by_freq}
\end{figure*}

% -------------------- Refs --------------------

\begin{acronym}
    \acro{SETI}{Search for Extraterrestrial Intelligence}
    \acro{ETA}{extraterrestrial agent}
    \acro{ATSAT}{Allen Telescope Survey Agnostic Technosignature pipeline}
    \acro{ETZ}{Earth Transit Zone}
    \acro{ASP}{Anti-Solar Point}
    \acro{ATA}{Allen Telescope Array}
    \acro{RFI}{radio frequency interference}
    \acro{SNR}{signal-to-noise ratio}
    % \acro{BLI}{Breakthrough Listen Initiative}
    % \acro{BL}{Breakthrough Listen}
    \acro{HCRO}{Hat Creek Radio Observatory}
    \acro{SOI}{signal of interest}
    \acro{SOIs}{signals of interest}
    \acro{BLADE}{Breakthrough Listen Accelerated DSP Engine}
    \acro{SI}{the SETI Institute}
\end{acronym}

\end{document}